\documentstyle[12pt,aaspp4]{article}

\newcommand{\etal}{et~al.}

\begin{document}

\title{
	The External Shear Acting on Gravitational Lens
	B~1422+231\altaffilmark{1}
}

\author{
	Tomislav Kundi\'c\altaffilmark{2},
	David W. Hogg\altaffilmark{2},
	Roger D. Blandford\altaffilmark{2},
	Judith G. Cohen\altaffilmark{3},
	Lori M. Lubin\altaffilmark{4} and 
	James E. Larkin\altaffilmark{5}
}

\altaffiltext{1}{Based on observations obtained at the W. M. Keck
Observatory, which is operated jointly by the California Institute of
Technology and the University of California}

\altaffiltext{2}{Theoretical Astrophysics, California Institute of
Technology, Mail Code 130-33, Pasadena, CA 91125; tomislav, hogg,
rdb@tapir.caltech.edu}

\altaffiltext{3}{Palomar Observatory, California Institute of
Technology, Mail Code 105-24, Pasadena, CA 91125;
jlc@astro.caltech.edu}

\altaffiltext{4}{The Observatories of the Carnegie Institution of
Washington, 813 Santa Barbara Street, Pasadena, CA 91101;
lml@ociw.edu}

\altaffiltext{5}{Palomar Observatory, California Institute of
Technology, Mail Code 320-47, Pasadena, CA 91125; currently at the
Department of Astronomy and Astrophysics, University of Chicago;
larkin@agrabah.uchicago.edu}

\begin{abstract}

In a number of multiply imaged quasar systems, a significant
contribution to the lensing potential is provided by groups and
clusters of galaxies associated with the primary lens.  As part of an
ongoing effort to gather observational data on these systems, we
present spectroscopy and near-infrared and optical photometry of
galaxies in the field of the quadruple lens system B~1422+231.  The
spectra show that the primary lens and five nearby galaxies belong to
a compact group at $z = 0.338$. The median projected radius of this
group is 35 $h^{-1}$ kpc and its velocity dispersion is $\sim$ 550 km
s$^{-1}$.  A straightforward application of the virial theorem yields
a group mass of $1.4 \times 10^{13} h^{-1} M_{\sun}$, which provides
sufficient external shear to produce the observed image configuration.
This data rules out a class of models and improves the system's
prospects for a measurement of the Hubble constant.

\end{abstract}


\section{Introduction \label{intro.sec}} 

One of the most exciting astrophysical applications of gravitational
lensing is the determination of the cosmological distance scale.  The
technique of measuring the Hubble constant in multiply imaged variable
sources was developed by Refsdal \markcite{REF64}(1964), who realized
that the time delay between variations of the individual images is
inversely proportional to $H_0$.  So far, two accurate time delays
have been reported in the literature, one in the double quasar
Q~0957+561 (Kundi\'c \etal\ \markcite{KUN96}1996) and the other in the
quadruple quasar PG~1115+080 (Schechter \etal\ \markcite{SCH97}1997),
yielding consistent values of the Hubble constant.  However, before
Refsdal's method is universally accepted as a competitor to the
traditional distance ladder approach, it has to be applied to a
statistically significant sample of systems with different image
morphologies and source and lens redshifts to verify the validity of
the lensing models (Blandford \& Kundi\'c \markcite{BK97}1997).  One
of the systems that shows good promise for extending the time delay
work is the quadruple quasar B~1422+231.

The gravitational lens system B~1422+231 ($V = 16.5$, $z = 3.62$) was
discovered by Patnaik \etal\ \markcite{PAT92a}(1992a) as part of the
JVAS radio survey (Patnaik \etal\ \markcite{PAT92b}1992b,
\markcite{PAT94}1994).  VLA and MERLIN maps of the system exhibited
four-image structure with three bright unresolved images (A, B and C)
opposite a much fainter image D.  The brighter components were shown
to have similar polarization properties and spectral indices, strongly
suggesting that they are lensed images of the same source.  Infrared
observations of the system (Lawrence \etal\ \markcite{LAW92}1992)
confirmed the relative positions and relative fluxes of the four
images.  The only disagreement between the radio and the optical data
is in the flux ratio of images A and B.  This is discussed in more
detail in \S\ref{model.sec}.  Optical imaging of the system revealed
the primary lens (Yee \& Ellingson \markcite{YE94}1994) and five
nearby galaxies (Remy \etal\ \markcite{REM93}1993, Bechtold \& Yee
\markcite{BY95}1995). A tentative redshift of the lensing galaxy was
obtained by Hammer \etal\ \markcite{HAM95}(1995).  Astrometry of the
system with the Faint Object Camera on the Hubble Space Telescope
(Impey \etal\ \markcite{IMP96}1996) confirmed the radio positions of
Patnaik \etal\ \markcite{PAT92a}(1992a) with 0.01$\arcsec$ accuracy.
FOS spectra of the four images were shown to be strikingly similar,
providing additional evidence for the lensing origin of the system.

Because this system shows evidence for variability in visual
luminosity (Hjorth \etal\ \markcite{HJO96}1996, Yee \& Bechtold
\markcite{YB96}1996, Turner \etal\ \markcite{TUR97}1997) and radio
morphology (Patnaik, private communication), it offers a good prospect
for measuring the Hubble constant.  For this reason it is extremely
important to constrain, as tightly as possible, the suite of
acceptable lens models.

In this paper we present spectroscopy and visual and near-infrared
imaging of the system with the goal of measuring the redshift of the
main lensing galaxy and its five neighbors, and estimating the
gravitational potential contributed by these galaxies based on their
mass-to-light ratios and the velocity dispersion of the group.

\section{Observations \label{obs.sec}}

\subsection{Optical and Near-Infrared Photometry}

$R$ band images of the B~1422+231 field were taken on 1995 July 30
with the Low Resolution Imaging Spectrograph (LRIS; Oke \etal\
\markcite{OKE95}1995) on the 10~m Keck telescope.  The scale on the
2048x2048 LRIS detector is 0.215 arcsec/pixel.  Exposures totaling 100
seconds were registered and combined to produce the final frame from
which relative positions and magnitudes of galaxies in the field were
measured. The FWHM of the point spread function in the combined image
was $0.75\arcsec$. A $2\arcmin \times 2\arcmin$ region of the frame is
shown in Fig.~\ref{Rimage.fig}.  The central object in the image
(marked B~1422+231) contains four quasar images and the main lensing
galaxy G1.  The remaining lens group galaxies (G2, G3, G4, G5 and G6)
are located southeast of the quasar.  Galaxy G7 was also detected in
near-infrared images (see below), but its spectrum is not available.
G8, G9 and G10 are emission line galaxies with redshifts $z_8 =
0.2355$, $z_9 = 0.4650$, and $z_{10} = 0.4555$. They are unrelated to
the lensing group (see below), but close enough to the line of sight
to the source to be incorporated perturbatively into any detailed lens
models.  The positions of all galaxies relative to image B and their
$R$ magnitudes measured in a 3$\arcsec$ circular aperture are listed
in Table~\ref{galdat.tab}.  Since we were unable to determine
accurately the position of image B, the coordinates of a nearby star
(S1) are also given in the table.  We estimate that the relative
positional error is $\lesssim 0.2\arcsec$, and the error in relative
magnitudes is $\lesssim 0.2$ mag.  The zero point of the $R$ magnitude
scale was determined by comparing our data with photometry of Remy
\etal\ \markcite{REM93}(1993).

Near-infrared $J$, $H$ and $K_s$ images of the field were taken in
March 1994 with the Near Infrared Camera (NIRC; Matthews \& Soifer
\markcite{MS94}1994) on the Keck telescope. The camera is equipped
with a $256 \times 256$ Santa Barbara Research Center InSb detector
with a scale of 0.15 arcsec/pixel, resulting in a field of view of
$38.4\arcsec \times 38.4\arcsec$.  The images were taken in a standard
``box 9'' pattern in which the telescope is moved roughly 5 arcsec
between exposure in a regular $3 \times 3$ grid.  Bright objects in
the frames are masked off and then the 9 images from each pattern are
medianed together to produce a sky and flat field frame.  Due to a
significant time varying background, a twilight or dome flat does not
provide adequate flat fielding.  Exposures in each band were
registered using bright objects in the field and were then averaged
together to yield the final images. The total exposure times were 120
s in $J$, 200 s in $H$, and 400 s in $K_s$ (Fig.~\ref{Kimage.fig}).
The FWHM of the point spread function was $0.7\arcsec$ in the combined
frames.  The zero point of the magnitude scale was determined by
repeated observations of 4 UKIRT faint standards (Casali \& Hawarden
\markcite{CH92}1992), yielding photometric uncertainties of roughly
0.025 mag for each of the infrared bands.  The resulting $J$, $H$ and
$K_s$ photometry of the lensing group galaxies is listed in
Table~\ref{galdat.tab}.  These magnitudes are quoted for a circular
aperture with a 3$\arcsec$ diameter.  Since the galaxies extend beyond
this radius, an aperture correction was applied before performing
mass-to-light ratio calculations in the following section.
Unfortunately, group galaxy G5 was not covered by the NIRC field of
view.

\subsection{Spectroscopy}

LRIS spectra of B~1422+231 were taken on 1995 July 30--31, 1997 March
2, and 1997 June 2.  On the first observing run (July 1995), a 300
line mm$^{-1}$ grating was used which provides spectral resolution of
2.47 \AA/pixel.  The second and third set of spectra (March and June
1997) were taken with a higher-resolution 600 line/mm grating yielding
1.25 \AA/pixel. A $1.0\arcsec$ slit was used throughout the observing
program.  The resulting FWHM of isolated sky lines was 10--15 \AA\ in
the low-resolution spectra and 4--5 \AA\ in the high-resolution
spectra.  Detailed observing parameters are listed in
Table~\ref{obslog.tab}.

The spectra of well separated group galaxies were reduced using
standard IRAF\footnotemark \footnotetext{IRAF is distributed by the
National Optical Astronomy Observatories, which are operated by the
Association of Universities for Research in Astronomy, Inc., under
cooperative agreement with the National Science Foundation.}
processing tasks. Wavelength solutions were derived from a fit to sky
lines on both sides of each spectrum, and independently confirmed with
arc lamp spectra taken immediately before and after the science
exposures.  The agreement between the two solutions was in all cases
better than 1 \AA. No attempt was made to flux-calibrate the spectra. 

\begin{figure}
\epsscale{0.9}
\plotone{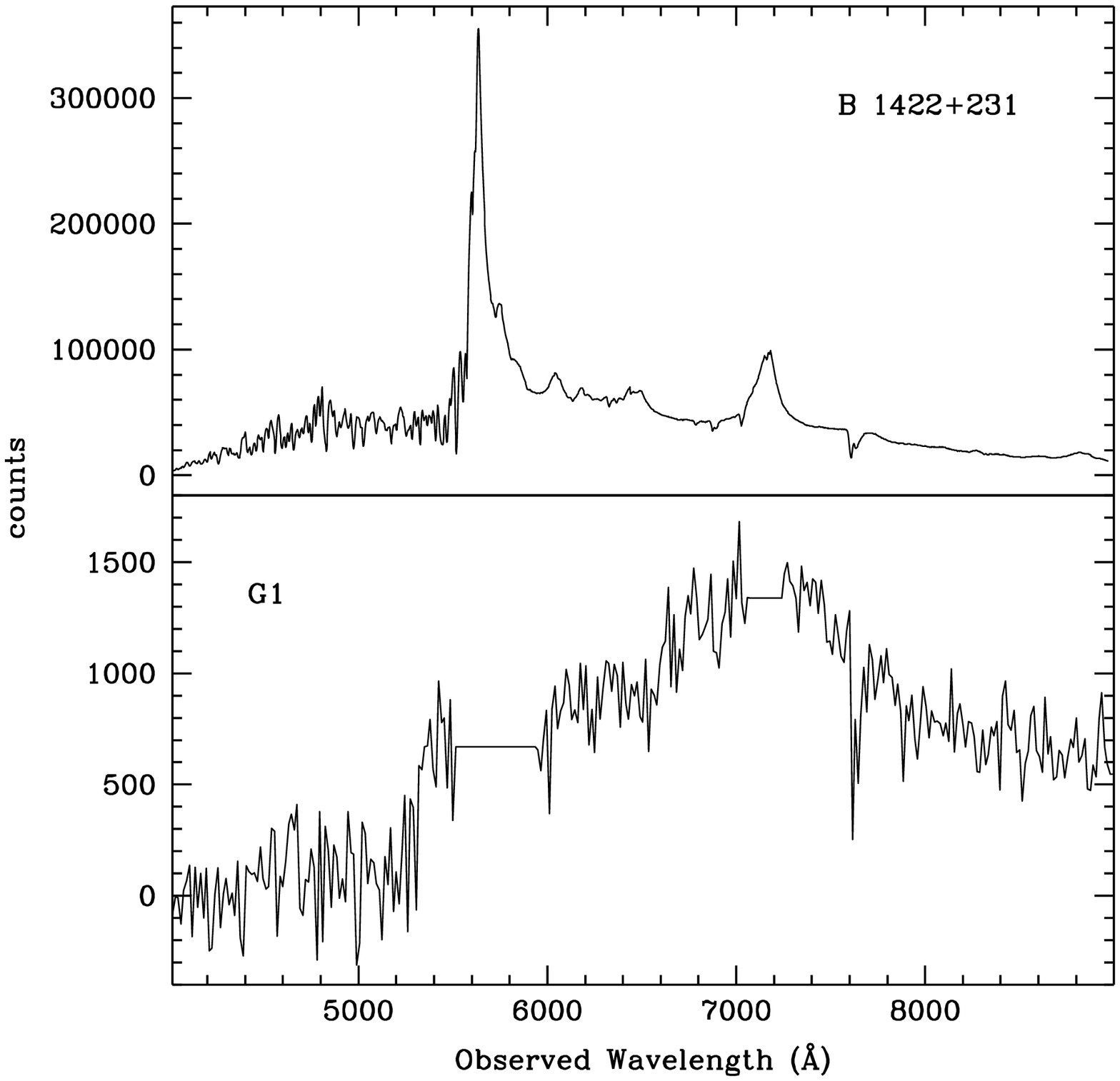}
\caption{Composite spectrum of B~1422+231 ({\em top}), and the primary
lens galaxy G1 ({\em bottom}). We have interpolated over the residuals
of broad quasar emission lines in the G1 spectrum and smoothed the
spectrum to 15\AA\ resolution. The redshift of G1 was determined from
the Ca H and K lines blueward of the 4000\AA\ break (observed at
5350\AA) and the Na D line observed at 7882 \AA.}
\label{qsoG.fig}
\end{figure}

The primary lensing galaxy G1 is within one arcsecond of the four
quasar images and almost 6 magnitudes fainter than their combined flux
(Impey \etal\ \markcite{IMP96}1996, Yee \& Ellingson
\markcite{YE94}1994). To remove the quasar light, it was thus
necessary to subtract a scaled quasar spectrum from the spectrum of G1
and interpolate over the remaining residuals of two strong quasar
emission lines (Ly~$\alpha$ at 5600 \AA\ and C~IV at 7200 \AA).  The
resulting galaxy spectrum is shown against the spectrum of the quasar
in Fig.~\ref{qsoG.fig}.  The quasar spectrum in this figure represents
the sum of unresolved image A, B, and C fluxes.

\begin{figure}
\epsscale{0.85}
\plotone{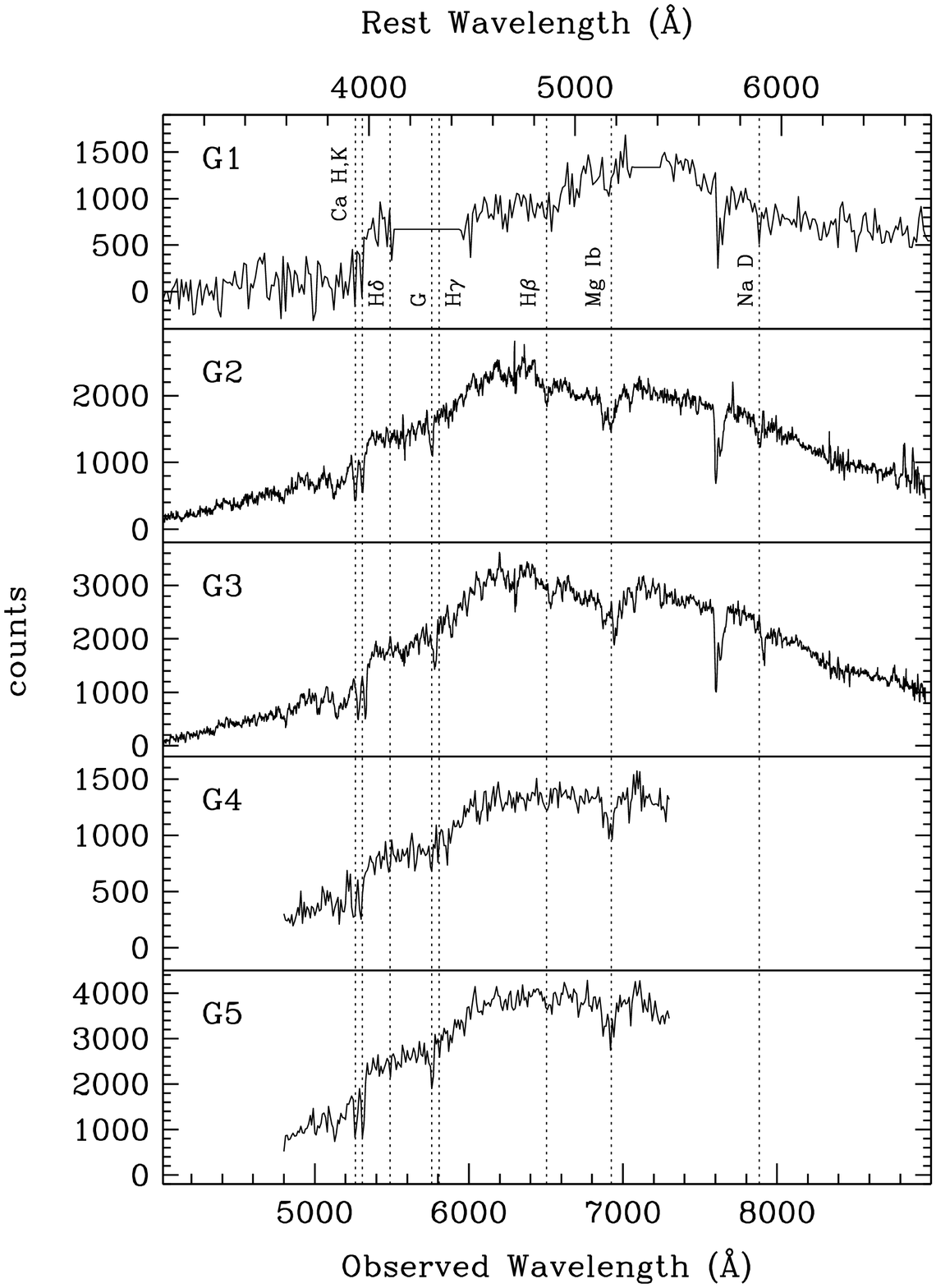}
\caption{The spectra of the primary lens G1 and group galaxies G2, G3,
G4 and G5 ({\em top to bottom}). In order to suppress noise, G1
spectrum is binned to 15\AA\ resolution, while G4 and G5 spectra are
binned to 10\AA\ resolution. High signal-to-noise G2 and G3 spectra
are shown at instrumental resolution. Strong spectral features are
identified with dotted lines assuming the group redshift of $z =
0.338$.  The same redshift is used for the rest wavelength scale on
the top of the figure. Note the redshift difference between G3 (the
brightest group member) and the other four galaxies.}
\label{5G.fig}
\end{figure}

We are unable to confirm the detection of weak [OII] and [OIII]
emission lines in the spectrum of G1 that lead Hammer \etal\
\markcite{HAM95}(1995) to the conclusion that the lensing galaxy
redshift is 0.647.  Instead, we find a break in the continuum at 5300
\AA\ and two absorption lines blueward of the break that are well
fitted by Ca H and K lines at $z = 0.3374$.  Detection of the Na D
line at 7882 \AA\ (top panel of Fig.~\ref{5G.fig}) provides further
confirmation of this redshift. 

Accurate redshifts for galaxies G2--G6 were determined by
cross-correlating their spectra with the templates of G and K giants
published by Jones \markcite{JON96}(1996) on the AAS CD-ROM 7
(Leitherer \etal\ \markcite{LEI97}1997).  The results of the
cross-correlation analysis are summarized in Table~\ref{galdat.tab}.
The errors listed in the table were calculated from the width of the
cross-correlation peak (Tonry \& Davis \markcite{TD79}1979) and an
estimate of the systematic error in wavelength calibration.  The
spectra of five lensing galaxies are shown in Fig.~\ref{5G.fig}.
Major absorption features are marked in these spectra with vertical
dotted lines.  A closer inspection of the figure reveals a significant
redshift difference ($\Delta z = 0.005$) between G3, the brightest
group galaxy, and the rest of the lensing group.  Galaxy G3 is thus a
large contributor to the high velocity dispersion of the group (see
Table~\ref{groups.tab}).

\section{A Comparison Between the PG~1115+080 and the B~1422+231
Lensing Groups \label{comp.sec}}

The close proximity of galaxies G1--G6 in redshift and angular
coordinates on the sky strongly suggests that they are members of a
gravitationally bound compact group of galaxies.  The rest-frame
line-of-sight velocity dispersion of this group, estimated with
$\sigma_v = c \sigma_z/(1 + z)$, is $550 \pm 50$ km s$^{-1}$.  The
median projected separation of group galaxies is 35 $h^{-1}$ kpc,
assuming $\Omega = 1$.  At lower redshifts similar groups have been
found by Hickson \markcite{HIC82}(1982) and studied in detail by
Hickson \etal\ \markcite{HIC92}(1992).  While the lensing group in the
B~1422+231 system has a higher velocity dispersion than is typical of
Hickson's compact groups (HCGs), it is still within the range of
Hickson's sample.  As we mentioned above, most of the velocity
dispersion is contributed by the brightest group galaxy G3, whose
rest-frame velocity is $\sim 1100$ km s$^{-1}$ higher than the median
velocity of the group.  This galaxy would be excluded by Hickson
\etal\ \markcite{HIC92}(1992) from the sample of accordant members,
which must have velocities within 1000 km s$^{-1}$ of the group
median.  Excluding G3, the rest-frame velocity dispersion of the group
would be only $240 \pm 60$ km s$^{-1}$.  However, because of its
magnitude and central location within the group we retain G3 in the
subsequent analysis.

There are at least three other QSO gravitational lens systems where
the primary lens is surrounded by a group or a cluster: the double
quasar Q~0957+561 (Walsh, Carswell, \& Weymann \markcite{WCW79}1979),
the quadruple quasar PG~1115+080 (Weymann \etal\
\markcite{WEY80}1980), and the recently discovered radio quad
MG~0751+2716 (Leh\'ar \etal\ \markcite{LEH97}1997).  In the
largest-separation confirmed lens system, Q~0957+561, the primary lens
is positioned near the center of a medium-rich cluster with the
velocity dispersion of $715 \pm 130$ km s$^{-1}$ (Angonin-Willaime,
Soucail, \& Vanderriest \markcite{ASV94}1994). Both PG~1115+080
(Kundi\'c \etal\ \markcite{KUN97}1997) and MG~0751+2716 are lensed by
compact groups of galaxies, although the redshifts of the MG~0751+2716
group galaxies are not yet available. 

In Table~\ref{groups.tab} we compare properties of the two lensing
groups that have been spectroscopically investigated at Keck:
PG~1115+080 and B~1422+231.  Both systems are similarly compact, but
the velocity dispersion is much higher in B~1422+231 as a result of
the large velocity difference between G3 and the rest of the group.
Assuming that these lensing groups are self-gravitating, their masses
can be estimated using the virial theorem (e.g. Heisler, Tremaine, \&
Bahcall \markcite{HTB85}1985):
\begin{equation}
M = \frac{3 \pi N}{2 G} \frac{\sum_i v_i^2}{\sum_{i < j} 1 / R_{\perp,
ij}} \quad ,
\label{virial.eq}
\end{equation}
where $v_i$ is the line-of-sight velocity of the $i$th galaxy relative
to the centroid of the group, and $R_{\perp, ij}$ is the projected
separation of galaxies $i$ and $j$.  The virial estimates, as well as
the light-weighted group centroids, are listed in
Table~\ref{groups.tab}.  These values should be treated with caution,
because of the small number of galaxies used to derive them.  The
assumptions inherent in the virial mass estimate (Eq.~\ref{virial.eq})
and its accuracy in reproducing the masses of simulated groups are
discussed by Heisler \etal\ \markcite{HTB85}(1985).  These authors
find that 75\% of mass estimates lie within 10$^{0.25}$ of the correct
value for groups with 5 members.

Frequent occurrence of groups and clusters near the primary lens is an
important clue for statistical studies of gravitational lensing.
Keeton, Kochanek, \& Seljak \markcite{KKS96}(1996, hereafter KKS96)
have convincingly shown that external shear is a more fundamental
parameter for the lensing models than the radial distribution of mass
in the primary lens.  External sources of shear thus warrant careful
observational scrutiny.

\section{Models \label{model.sec}}

Shortly after discovery of B~1422+231, lens models of this system were
constructed by Hogg \& Blandford (\markcite{HB94}1994, hereafter HB94)
and by Kormann, Schneider, \& Bartelmann (\markcite{KSB94}1994).  It
was realized by both groups that a substantial ellipticity in the
potential is required to successfully reproduce the observed
morphology of the system.  However, the two sets of models disagree on
the relative importance of external shear and internal ellipticity in
breaking the circular symmetry of the lens potential.  HB94 rely on
strong external shear from two nearby bright galaxies, G2 and G3,
while Kormann \etal\ make G1 highly elliptical.  Subsequent {\em HST}
observations of the lens (Impey \etal\ \markcite{IMP96}1996) showed
that its optical axis ratio $a/b = 1.37 \pm 0.14$ is smaller than the
ratio predicted by the Kormann \etal\ \markcite{KSB94}(1994) models,
$1.68 < a/b < 2.86$.  The observations can only be reconciled with the
models if the dark matter distribution in the lens is significantly
flatter than its optical isophotes.

More recently, KKS96 modeled B~1422+231 as part of their investigation
of shear and ellipticity in gravitational lenses.  A new observational
input for these models was a precise position of the primary lens G1
provided by the {\em HST} imaging (Impey \etal\
\markcite{IMP96}1996). KKS96 distinguish between two classes of
models: single-shear and multiple-shear.  Single-shear models feature
only one source of ellipticity which is provided either by the primary
lens or by an external shear field.  In the two-shear models, there
are two independent sources of ellipticity with major axes that are
generally misaligned.  Based on previous work (Kochanek
\markcite{KOC91}1991, Wambsganss \& Paczy\'nski \markcite{WP94}1994)
and their own models, KKS96 argue that adding additional parameters to
the radial distribution of a model generally does little to improve
the fit.  On the other hand, KKS96 show that including additional
sources of shear often results in dramatic improvements.
Unfortunately, this is not true for B~1422+231.  The $\chi^2$ of the
one-shear model, in which the primary lens is modeled as a singular
isothermal sphere perturbed by a quadrupole shear potential, is 40.3
for 6 degrees of freedom and 9 parameters.  The corresponding
two-shear model, consisting of a singular isothermal ellipsoid in an
external shear field, has a $\chi^2$ of 33.7 and 4 degrees of freedom
(KKS96).  The justification of adding new parameters to the model can
be assessed with a standard F-test (e.g. Lupton \markcite{LUP93}1993).
According to this test, the probability of $\chi^2$ decreasing from
40.3 to 33.7 when two {\em arbitrary} parameters are added to the
model is 76\%, suggesting that the second shear does not significantly
reduce the $\chi^2$.  Normally, the extra parameters would be accepted
only if the F-test probability were smaller than 5\%.  Another reason
to doubt the importance of the second shear term is the alignment of
the two axes in the two-shear model: the angle between them is only
4$^\circ$ (KKS96).  We are left with the conclusion that the best
current model of the system consists of a singular isothermal sphere
in the external shear field of the group, which is essentially the
original HB94 model.

There are, however, two potential problems with the HB94 model: it
requires large external shear from a massive perturber, and it fails
to reproduce the flux ratio of images A and B at radio wavelengths.
We consider these two problems in turn.

In the previous section, we estimated the lensing group mass using the
virial theorem.  We can divide this value with the combined $K_s$
luminosity of group members, and compare the resulting near-infrared
mass-to-light ratio of the lensing group with optical ratios of local
groups.  The total near-infrared luminosity of the lensing group
galaxies, corrected by $-0.3$ mag to account for the light outside of
the $3\arcsec$ aperture, is $K_s=14.9$.  At the redshift of $z=0.338$,
this corresponds to an absolute magnitude of $M_{Ks}=-25.5 + 5 \log
h$, where $H_0 \equiv 100 \, h$ km s$^{-1}$ Mpc$^{-1}$ and $(\Omega,
\Lambda) = (1, 0)$.  We applied a $K$-correction of $-0.2$ mag
(Poggianti \markcite{POG97}1997) and neglected the evolutionary
corrections and the difference between the $K_s$ and $K$ bands.  Using
the absolute magnitude of the Sun, $M_{K\sun} = +3.4$ (Allen
\markcite{ALL73}1973), we find that the near-infrared mass-to-light
ratio of the group is 40$h$ in solar units.  In a large sample of
nearby groups Ramella \etal\ \markcite{RPG97}(1997) find that their
blue mass-to-light ratios range between 160$h$ and 330$h$.  Taking
into account that the mass-to-light ratio of an old stellar population
is 6-8 times higher in blue than it is in $K$ (Worthey 1994), we
conclude that the B~1422+231 lensing group fits well within the
Ramella \etal\ sample.

For the purpose of lens modeling, we can directly use the measured
velocity dispersion of the group, but we have to make certain
assumptions about the mass profile.  If we model the group as a
singular isothermal sphere, its velocity dispersion of 550 km s$^{-1}$
acting at an angular distance of $\sim 14$~arcsec will produce the
dimensionless shear of $\gamma = 0.23$ at the position of the lens.
This shear is consistent with the value expected in the HB94 model and
the related one-shear KKS96 model.  It is also worth noting that group
galaxies G2--G6 are so close to the line of sight to the QSO that they
undoubtedly contribute to the convergence (dimensionless surface mass
density) as well as the shear.  If the lensing group has singular
isothermal sphere profile, the convergence $\kappa$ is equal to the
shear $\gamma$.  A uniform convergence term in the model does not
affect the observed image positions and relative fluxes; however, it
will affect the inferred distance to the lens when the time delays are
measured (Falco, Gorenstein, \& Shapiro \markcite{FGS85}1985).

We have accounted for the presence of a massive perturber, but the
fact that the HB94 model fails to reproduce the radio flux ratio of
images A and B remains a serious problem.  In the refined model of
HB94 the relative magnification of these two images is 0.77, many
standard deviations away from the radio measurement of $0.98 \pm 0.02$
(Patnaik \etal\ \markcite{PAT92a}1992a).  However, optical and
near-infrared measurements have consistently produced lower flux
ratios clustering around 0.8 (Lawrence \etal\ \markcite{LAW92}1992,
Remy \etal\ \markcite{REM93}1993, Yee \& Ellingson
\markcite{YE94}1994, Hammer \etal\ \markcite{HAM95}1995, Akujor \etal\
\markcite{AKU96}1996, Impey \etal\ \markcite{IMP96}1996, Yee \&
Bechtold \markcite{YB96}1996).  It is difficult to explain the
difference between the radio and the optical as a result of absorption
along the line of sight to the quasar because the flux ratio is nearly
identical in all bands between $U$ and $K_s$.  It is also unlikely
that it can be explained by microlensing since the ratio has not
changed over at least a year (Yee \& Bechtold \markcite{YB96}1996) and
even longer if data taken by different observers are included.
Regardless of which ratio represents the true relative magnification
of the two images, it is important that we explain the observed
discrepancy, as it may have profound consequences for models of other
gravitational lens systems.

\section{Discussion}

Based on Keck spectroscopy of the gravitational lens system
B~1422+231, we conclude that the primary lens and four nearby objects
belong to a compact group of galaxies at $z_d = 0.338$.  The virial
mass of this group, inferred from its median projected radius of 35
$h^{-1}$ kpc and its velocity dispersion of $\sim$ 550 km s$^{-1}$,
amounts to $1.4 \times 10^{13} \, h^{-1} \, M_{\sun}$.  This
concentration of mass, located some $14\arcsec$ from the quasar
images, is sufficient to produce the observed image configuration in a
simple lens model consisting of a singular isothermal sphere in a
strong external shear field.  The same model successfully accounts for
the optical and near-infrared flux ratios of the images, although it
remains inconsistent with the relative magnification of images A and B
in radio. An accurate lens model for this system is of great
importance because photometric monitoring shows the source to be
intrinsically variable on the timescales comparable to the predicted
time delay between the triplet ABC and image D (Hjorth \etal\
\markcite{HJO96}1996, Yee \& Bechtold \markcite{YB96}1996, Turner
\etal\ \markcite{TUR97}1997).  Using the lens redshift of $z_d =
0.338$, the HB94 model predicts this time delay to be about 15
$h^{-1}$ days.  The delays between images A, B and C are all smaller
than a day -- too short to measure at optical and radio wavelengths.

By trimming down model space and reducing the uncertainty in predicted
time delays, the observations presented here improve the chances of
using this system to measure the Hubble constant.  Together with the
spectroscopic observations of PG~1115+080 (Kundi\'c \etal\
\markcite{KUN97}1997), they also suggest that a number of
gravitational lens systems are influenced by significant external
shear.  The external shear provided by galaxies associated with the
primary lens naturally explains why many lens models require
potentials that are more elliptical than the images of the primary
lensing galaxy would suggest.  External shear could also be
responsible for the large number of quadruple lenses relative to the
doubles (King \etal\ \markcite{KIN96}1996).  A spectroscopic survey of
the fields around other gravitational lenses would thus provide
valuable observational input for estimates of external shear in these
systems.

\acknowledgements

We are grateful to the W. M. Keck Foundation for the vision to fund
the construction of the W. M. Keck Observatory.  We thank Charles
Lawrence, Keith Matthews, Gerry Neugebauer and Tom Soifer for making
their unpublished near-infrared data available.  We benefited from
helpful conversations with Sangeeta Malhotra and Michael Pahre.
Financial support was provided by NSF grants AST~92--23370 and
AST~95--29170, and by NASA grant NAG~5-3834.  This research made use
of NASA's Astrophysics Data System Abstract Service.

\clearpage

\begin{deluxetable}{crrccccccc}
\tablecolumns{10}
\tablewidth{0pc}
\tablecaption{Lensing Galaxy Positions, Magnitudes and Redshifts}
\tablehead{
   \colhead{Galaxy} & \colhead{$-\Delta \alpha$ ($\arcsec$)} &
   \colhead{$\Delta \delta$ ($\arcsec$)} &
   \colhead{$V$\tablenotemark{1}} & \colhead{$R$} &
   \colhead{$i$\tablenotemark{1}} &
   \colhead{$J$} & \colhead{$H$} & \colhead{$Ks$} &
   \colhead{Redshift}
}
\startdata
G1 & $ -0.7\phn\phn$ & $ -0.6\phn\phn$ & 21.5\tablenotemark{2} &
   $r=21.8$\tablenotemark{3} & \nodata &
   \nodata & \nodata & \nodata & $0.3374 \pm 0.0005$ \nl
G2 & $ -9.4\phn\phn$ & $ -4.9\phn\phn$ & 20.4    & 19.9 & 19.1    &
   17.1    & 17.9    & 16.4    & $0.3379 \pm 0.0005$ \nl
G3 & $ -3.9\phn\phn$ & $ -7.2\phn\phn$ & 20.0    & 19.6 & 18.8    &
   16.7    & 17.5    & 16.0    & $0.3427 \pm 0.0005$ \nl
G4 & $ -7.5\phn\phn$ & $-10.4\phn\phn$ & 21.6    & 21.0 & 20.5    &
   18.4    & 19.0    & 17.6    & $0.3366 \pm 0.0005$ \nl
G5 & $-17.4\phn\phn$ & $-28.8\phn\phn$ & 20.3    & 20.0 & 19.4    &
   \nodata & \nodata & \nodata & $0.3384 \pm 0.0005$ \nl
G6 & $  4.4\phn\phn$ & $ -7.4\phn\phn$ & \nodata & 22.2 & \nodata &
   20.2    & 20.3    & 19.9    & $0.3357 \pm 0.0005$ \nl
G7 & $ 16.7\phn\phn$ & $-12.0\phn\phn$ & \nodata & 23.0 & \nodata &
   18.9    & \nodata & 18.4    & \nodata             \nl
G8 & $ -2.1\phn\phn$ & $ 12.9\phn\phn$ & \nodata & 22.4 & \nodata &
   20.2    & 20.7    & 20.0    & $0.2355 \pm 0.0005$ \nl
G9 & $ 25.5\phn\phn$ & $ 35.0\phn\phn$ & \nodata & 21.7 & \nodata &
   \nodata & \nodata & \nodata & $0.4650 \pm 0.0005$ \nl
G10& $ 16.6\phn\phn$ & $-41.5\phn\phn$ & \nodata & 22.2 & \nodata &
   \nodata & \nodata & \nodata & $0.4055 \pm 0.0005$ \nl
S1 & $  6.9\phn\phn$ & $-15.9\phn\phn$ & 18.1    & 17.4 & 17.1    &
   15.7    & \nodata & 15.6    & \nodata             \nl 
\enddata
\tablecomments{(1) $V$ and $i$ magnitudes are taken from Remy \etal\
\markcite{REM93}(1993); (2) $V$ magnitude of G1 is taken from Impey
\etal\ \markcite{IMP96}(1996); (3) Gunn $r$ magnitude of G1 is adopted
from Yee \& Ellingson \markcite{YE94}(1994).}
\label{galdat.tab}
\end{deluxetable}

\begin{deluxetable}{clccccccc}
\footnotesize
\tablecolumns{9}
\tablewidth{0pc}
\tablecaption{Observing Parameters}
\tablehead{
   \colhead{Exposure} & \colhead{Object} & \colhead{UT} &
   \colhead{UT} & \colhead{Airmass} & \colhead{PA} & 
   \colhead{Exposure} & \colhead{Grating} &
   \colhead{Wavelength} \nl
   \colhead{Number} & & \colhead{Date} & \colhead{Time} & &
   \colhead{($\arcdeg$E of N)} &
   \colhead{Time ($s$)} & \colhead{lines/mm} & \colhead{Range (\AA)} 
}
\startdata
1 & G2, G3      & 1995 Jul 30 & 05:55 & 1.09 &  69 & 1800 & 300 & 3960-8970 \nl
2 & G2, G3      & 1995 Jul 30 & 06:27 & 1.16 &  69 & 1200 & 300 & 3960-8970 \nl
3 & QSO, G1, G9 & 1995 Jul 31 & 06:14 & 1.14 & 315 & 1000 & 300 & 3960-8970 \nl
4 & QSO, G1, G9 & 1995 Jul 31 & 06:32 & 1.19 & 315 & 1000 & 300 & 3960-8970 \nl
5 & QSO, G1     & 1995 Jul 31 & 07:07 & 1.32 & 315 & 1000 & 300 & 3960-8970 \nl
6 & G4, G5      & 1997 Mar  2 & 14:17 & 1.00 & 332 & 1200 & 600 & 4770-7310 \nl
7 & G5          & 1997 Mar  2 & 14:52 & 1.02 & 327 & 1200 & 600 & 4770-7310 \nl
8 & G6, G8, G10 & 1997 Jun  2 & 09:12 & 1.04 &  18 & 1600 & 600 & 4260-6780 \nl
9 & G6, G8, G10 & 1997 Jun  2 & 09:41 & 1.09 &  18 & 1000 & 600 & 4260-6780 \nl
10& G6, G8, G10 & 1997 Jun  2 & 09:59 & 1.13 &  18 & 1800 & 600 & 5640-8190 \nl
\enddata
\label{obslog.tab}
\end{deluxetable}

\begin{deluxetable}{lcc}
\tablecolumns{3}
\tablewidth{0pc}
\tablecaption{Lensing Group Properties: PG~1115+080 and B~1422+231}
\tablehead{
   \colhead{Gravitational Lens System} &
   \colhead{PG~1115+080} &
   \colhead{B~1422+231} 
}
\startdata
Number of Galaxies                                &
   4                              & 6                       \nl
Group Redshift                                    &
   0.311                          & 0.338                   \nl
Velocity Dispersion (km s$^{-1}$)                 &
   $270\pm70$                     & $550\pm50$              \nl
Median Projected Galaxy Separation ($h^{-1}$ kpc) &
   35                             & 35                      \nl
Virial Mass ($h^{-1} \,M_{\sun}$)                 &
   $4.6 \times 10^{12}$           & $1.4 \times 10^{13}$    \nl
Group Centroid ($d$, $\theta$)\tablenotemark{1}   &
   ($17\arcsec$, $-121^\circ$) & ($14\arcsec$, $146^\circ$) \nl
\enddata
\tablecomments{(1) The luminosity-weighted center of the group is
specified relative to image C in PG~1115+080 and relative to image B
in B~1422+231.  The distance of the centroid is quoted in
arcseconds and its position angle is measured north through east.}
\label{groups.tab}
\end{deluxetable}

\clearpage

\figcaption{$R$ band image of the $2\arcmin \times 2\arcmin$ field
surrounding the gravitational lens system B~1422+231. North is up and
east is to the left. The central bright object contains four
unresolved quasar images and the primary lens galaxy G1. The remaining
lensing group galaxies G2--G6 are located southeast of the quasar. G7
has not been studied spectroscopically, while G8, G9 and G10 are
emission line galaxies with redshifts 0.2355, 0.4650 and
0.4055. Positional reference star S1 is also marked in the figure.
\label{Rimage.fig}}

\figcaption{$K_s$ band image of B~1422+231 in $0.7\arcsec$ seeing. The
field is 38.4$\arcsec$ on the side with north up and east to the left.
Object names are same as in Fig.~\ref{Rimage.fig}.  The location of
the luminosity-weighted centroid of the group is roughly coincident
with G4.
\label{Kimage.fig}}

\end{document}